\documentclass[twocolumn,trackchanges,linenumbers]{aastex631}
\usepackage{amsmath,booktabs,amssymb}
\usepackage{multirow}
\usepackage[version=4]{mhchem}
\usepackage{tabularx}
\usepackage{threeparttable}
\usepackage{booktabs}
\usepackage[version=4]{mhchem}
\usepackage{subfigure}
\usepackage{comment}
\begin{document}

\title{Rotational Spectrum and First Interstellar Detection of 2-Methoxyethanol Using ALMA Observations of NGC 6334I} 

\author[0000-0001-5020-5774]{Zachary T. P. Fried}
\affiliation{Department of Chemistry, Massachusetts Institute of Technology, Cambridge, MA 02139, USA}
\author{Samer J. El-Abd}
\affiliation{Department of Astronomy, University of Virginia, Charlottesville, VA 22904, USA}
\affiliation{National Radio Astronomy Observatory, Charlottesville, VA 22903, USA}
\author[0000-0002-5849-8433]{Brian M. Hays}
\affiliation{Univ. Lille, CNRS, UMR 8523 - PhLAM - Physique des Lasers Atomes et Mol\'{e}cules, F-59000 Lille, France}
\author[0000-0002-0332-2641]{Gabi Wenzel}
\affiliation{Department of Chemistry, Massachusetts Institute of Technology, Cambridge, MA 02139, USA}
\author{Alex N. Byrne}
\affiliation{Department of Chemistry, Massachusetts Institute of Technology, Cambridge, MA 02139, USA}
\author{Laurent Margul\`es}
\affiliation{Univ. Lille, CNRS, UMR 8523 - PhLAM - Physique des Lasers Atomes et Mol\'{e}cules, F-59000 Lille, France}
\author{Roman A. Motiyenko}
\affiliation{Univ. Lille, CNRS, UMR 8523 - PhLAM - Physique des Lasers Atomes et Mol\'{e}cules, F-59000 Lille, France}
\author{Steven T. Shipman}
\affiliation{BrightSpec, Inc., Charlottesville, VA 22903, USA}
\author{Maria P. Horne}
\affiliation{Department of Chemistry, New College of Florida, Sarasota, FL 34243, USA}
\author{Jes K. J{\o}rgensen}
\affiliation{Centre for Star and Planet Formation, Niels Bohr Institute \& Natural History Museum of Denmark, University of Copenhagen,
\O{}ster Voldgade 5–7, 1350 Copenhagen K., Denmark}
\author{Crystal L. Brogan}
\affiliation{National Radio Astronomy Observatory, Charlottesville, VA 22903, USA}
\author{Todd R. Hunter}
\affiliation{National Radio Astronomy Observatory, Charlottesville, VA 22903, USA}
\author{Anthony J. Remijan}
\affiliation{National Radio Astronomy Observatory, Charlottesville, VA 22903, USA}
\author[0000-0002-6667-7773]{Andrew Lipnicky}
\affiliation{National Radio Astronomy Observatory, Charlottesville, VA 22903, USA}
\author{Ryan A. Loomis}
\affiliation{National Radio Astronomy Observatory, Charlottesville, VA 22903, USA}
\author[0000-0003-1254-4817]{Brett A. McGuire}
\affiliation{Department of Chemistry, Massachusetts Institute of Technology, Cambridge, MA 02139, USA}
\affiliation{National Radio Astronomy Observatory, Charlottesville, VA 22903, USA}

\correspondingauthor{Zachary T. P. Fried, Brett A. McGuire}
\email{zfried@mit.edu, brettmc@mit.edu}

\begin{abstract}
\nolinenumbers
We use both chirped-pulse Fourier transform and frequency modulated absorption spectroscopy to study the rotational spectrum of 2-methoxyethanol (\ce{CH3OCH2CH2OH}) in several frequency regions ranging from $8.7-500\,\mathrm{GHz}$. The resulting rotational parameters permitted a search for this molecule in Atacama Large Millimeter/submillimeter Array
(ALMA) observations toward the massive protocluster NGC 6334I as well as source B of the low-mass protostellar system IRAS 16293-2422. 25 rotational transitions are observed in the ALMA Band 4 data toward NGC 6334I, resulting in the first interstellar detection of 2-methoxyethanol. A column density of $1.3_{-0.9}^{+1.4} \times 10^{17}$ cm$^{-2}$ is derived at an excitation temperature of $143_{-39}^{+31}$ K. However, molecular signal is not observed in the Band 7 data toward IRAS 16293-2422B and an upper limit column density of $2.5 \times 10^{15}$\,cm$^{-2}$  is determined. Various possible formation pathways--including radical recombination and insertion reactions--are discussed. We also investigate physical differences between the two interstellar sources that could result in the observed abundance variations.

\end{abstract}

\keywords{}
\nolinenumbers
\section{Introduction}

\label{sec:intro}
Several molecules that include the methoxy (\ce{OCH3}) functional group have been detected in the interstellar medium, including dimethyl ether \citep{dme}, methyl formate \citep{mf}, ethyl methyl ether \citep{eme_orion}, and methoxymethanol \citep{methoxy_detection}. In fact, this radical itself has been detected in radio telescope observations \citep{methoxy_radical}. While methoxylated molecules have been predominately observed in warmer regions of space surrounding forming stars, both dimethyl ether and methyl formate have also been detected in several cold sources, such as dark clouds and prestellar cores \citep{bac12, methoxy_radical,jim16, taq17,som18,agu19,agu21}. 

Recently, as an attempt to predict strong molecular candidates for detection in the Class 0 protostellar source IRAS 16293-2422B (hereafter referred to as IRAS 16293B), \citet{fried} trained a machine learning model to map the molecular identities of the detected species to their observed column densities. The trained regressors could then be used to predict the column density of any other molecule in this region of interstellar space. Due to the large column densities of methoxylated molecules observed in this source, a few of the highest predicted abundance species also contained the methoxy functional group. One especially notable result was the high predicted column density of 2-methoxyethanol (\ce{CH3OCH2CH2OH}). The notable column density prediction for this molecule along with the simulated rotational parameters suggested that this molecule was a strong candidate for detection. 

Two groups have previously investigated the rotational spectrum of this molecule, both using Stark modulated spectroscopic techniques. The first study was conducted by \citet{buckley}. The authors measured 38 lines of the ground vibrational state of the most stable conformer in the frequency range of $8-26.5\,$GHz and derived values of the $A$, $B$, and $C$ rotational constants. They also determined the dipole moments to be 2.03 D for $\mu_a$, 1.15 D for $\mu_b$ and $\sim$0.25 D for $\mu_c$. Following this, \citet{caminati} further extended the rotational spectrum to around 40\,GHz. They measured ground state rotational transitions up to $J = 70$ and included all of the quartic distortion constants along with two sextic distortion constants in their fit. The rotational spectrum of this molecule has not been measured beyond $\sim$40 GHz. This, however, is necessary for observations in the star-forming regions where \citet{fried} and others predict potential detection, as at the warmer excitation temperatures (ranging to several hundred Kelvin in these sources), the strongest transitions of these species arise in the millimeter (mm) and submillimeter (sub-mm) regimes.

In this paper, we extend the rotational spectrum of the ground vibrational state of 2-methoxyethanol into the mm and sub-mm regions. Using these newly acquired data, we report the first interstellar detection of this molecule using Atacama Large Millimeter/submillimeter Array (ALMA) data toward the massive protocluster NGC 6334I \citep{hunter}. A search for this molecule was also conducted in the ALMA data toward IRAS 16293B; however, clear molecular signal was not observed. Instead, an upper limit of the column density was derived. 

\section{Quantum Chemical Calculations}
\begin{figure}[h!]

\begin{center}
\includegraphics[width=\columnwidth]{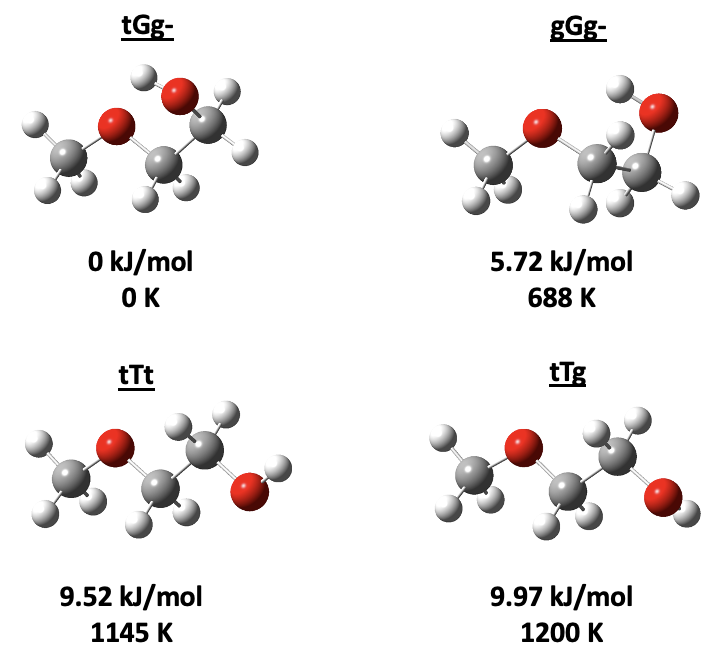}
\caption{
The structures and relative zero-point corrected energies (in kJ/mol and kT-equivalent temperature) of the four most stable conformers of 2-methoxyethanol. See text for computational details.}
\label{fig:conformers}
\end{center}
\end{figure}

The highly saturated nature of 2-methoxyethanol lends the possibility of various stable conformers. In a computational study of the thermal degradation of this molecule, \citet{conformers} computed the energies and molecular geometries of its 12 conformers. Using their reported geometries as inputs (which were optimized using the complete basis set CBS-QB3 ab initio method), we re-ran the geometry and energetic calculations in order to obtain the vibrational level frequencies and relative conformer energies. The geometries and harmonic vibrational frequencies of these molecules were calculated using Gaussian 16 \citep{g16} with the MP2 level of theory and aug-cc-pVTZ basis set. Tight convergence criteria was used for the geometry optimizations. Single point energy calculations were then conducted on the MP2 optimized geometries using the CCSD(T)/aug-cc-pVTZ functional and basis set. MP2/aug-cc-pVTZ zero point corrections were added to the single point energy calculations. The geometry and relative zero-point corrected energies of these conformers are shown in Figure~\ref{fig:conformers}. The tGg- conformer was calculated to be 5.72\,kJ/mol more stable than any other conformational arrangement of this molecule. The two lowest energy conformers, tGg- and gGg-, are stabilized in part due to the intramolecular hydrogen bonding interaction between the hydroxyl OH group and the methoxy oxygen lone pair electrons.

The lowest lying computed vibrationally excited states of the tGg- conformer are noted in Table~\ref{table:vibrations}. Also listed in this table are the energy values of these vibrations that were experimentally determined by \citet{caminati}, who measured several quanta of the states corresponding to the C-O torsion and C-C torsion. There are four states with a kT-equivalent temperature less than $\sim$400\,K above the ground vibrational state, thus resulting in a notable population of these excited levels at room temperature, which was the experimental temperature of the $150-500\,$GHz spectrum. 

\begin{table*}
\centering
\caption{Calculated harmonic vibrational frequencies, energies, and kT-equivalent temperatures of the five lowest lying vibrational states along with their physical description. Energies are reported relative to the ground vibrational state. Calculations were conducted using Gaussian 16 with the MP2 level of theory and the aug-cc-pVTZ basis set.}
\begin{tabular}{c c c c c}
\toprule
Caminati et al.    &   \multicolumn{3}{c}{Calculated} & \\
Frequency (cm$^{-1}$) &  Frequency (cm$^{-1}$) &  Energy (kJ/mol) & Energy (K) & Description \\
\midrule
 89 & 94.3& 1.13 & 135.7& O-C Torsion \\
 152 & 143.9 & 1.72 &   207.0& C-C Torsion \\
 221& 233.1& 2.79 & 335.3&  Methyl Torsion \\
 -- & 279.9 & 3.35 & 402.6& In-Plane Wag\\
 -- & 370.2 & 4.43  & 532.6& Symmetric In-Plane Wag  \\
\bottomrule
\end{tabular}
\label{table:vibrations}
\end{table*}

\citet{lee_bayesian} conducted a statistical analysis of the rotational constant calculations of various low-cost functional and basis set combinations. They determined that the $\omega$B97X-D/6-31+G(d) level of theory and basis set provided a strong balance between calculation uncertainty and efficiency. We therefore computed the $A$, $B$, and $C$ rotational constants through Gaussian 16 using these methods. These computed constants are reported in Table \ref{table:parameters}. Using the calculated values, simulated rotational spectra were generated using SPCAT in Pickett’s CALPGM suite of programs \citep{spcat}. All of the computational outputs are available in the Supplementary Information. 

\label{sec:calc}

\section{Spectroscopic Analysis and Results}
\label{sec:exp_results}

The molecular sample was purchased with 99.8\,\% purity from Sigma Aldrich and was used as-is without additional purification. The spectrum reported in this work was collected using three different spectrometers. The mm and sub-mm spectrum was predominately measured using the Lille spectrometer \citep{lille} with fast-scan mode \citep{fast_scan} in three frequency regions: $150-220\,$GHz, $230-330\,$GHz, and $370-500\,$GHz. The $68-85\,$GHz spectrum was recorded with a newly updated chirped-pulse Fourier transform (CP-FT) mm-wave spectrometer at MIT. Finally, centimeter-wave data from $8.7-26.5\,$GHz was collected at the New College of Florida using a waveguide-based CP-FT spectrometer \citep{ncf}. A thorough description of each experimental setup is provided in Appendix~\ref{apendix_a}.

As mentioned in Section~\ref{sec:intro}, the rotational spectrum of this molecule was previously studied from $\sim8-40\,$GHz. However, the microwave spectrum was re-measured in this work since newer chirped-pulse Fourier transform spectrometers are generally more sensitive than the Stark-modulated spectrometers employed in the earlier studies. 

The assignment of the ground vibrational state was initiated using the simulated spectrum that was generated from the rotational constants reported by \citet{buckley}. This allowed for the identification of several strong low-$K_a$ a-type transitions and then subsequent refinement of the constants and fit. In all, a total of 2,172 distinct lines were identified and measured. Including blended lines, this corresponded to 3,396 transitions included in the fit. These were analyzed with a least-squares fitting procedure using ASFIT \citep{asfit}. The transitions range from $J^{\prime\prime} = 2-102$ and $K_a^{\prime\prime} = 0-32$. Owing to a very strong a-type dipole moment, the majority of the measured lines were a-type transitions and numerous strong R-branch series of a-type lines were identified. Additionally, none of the a-type transitions displayed any A-E splitting that is characteristic of species containing methyl rotors. While the b-type lines were generally much weaker, several Q-branch b-type series ranging from $K_a^{\prime\prime} = 20-25$ were reliably identified and fit. No c-type transitions were observed since the dipole moment along this axis is very limited.

 Because the most stable conformer is near the prolate symmetric top limit ($\kappa$ = -0.948), the fit was conducted using both the A and S reductions with the $I^r$ representation. The constants generated by both reductions, those reported by \citet{caminati} and \citet{buckley}, as well as the parameters from quantum chemical calculations are displayed in Table~\ref{table:parameters}.

\begin{table*}
    \caption{Rotational parameters of the ground vibrational state of the tGg- 2-methoxyethanol conformer from this work along with the previous rotational studies of this molecule and theoretical calculations. }
    \label{table:parameters}
    \fontsize{7pt}{7pt}\selectfont
    \begin{threeparttable}
    \centering
    \begin{tabular}{llllllll}
    \hline
    \multicolumn{6}{l}{A Reduction} & \multicolumn{2}{l}{S Reduction}\\
    \cline{1-5}
    \cline{7-8}
     Parameter & Current Work & \cite{caminati}& \cite{buckley} & Theory\tnote{a} & & Parameter & Current Work  \\
    \hline
    $A$ (MHz) & 12982.3438(23)& 12982.398 & 12982.35& 12998.967& & $A$ (MHz) & 12982.3436(24) \\
    $B$ (MHz) & 2742.50569(10) & 2742.502& 2742.48& 2755.126& & $B$ (MHz) & 2742.49749(10)\\
    $C$ (MHz) & 2468.107109(94)& 2468.1039& 2468.10& 2476.547& & $C$ (MHz) & 2468.115202(99)\\
    \\
    $\Delta_{J}$ (kHz) & 1.423791(25)& 1.396 & & & & $D_{J}$ (kHz) & 1.396709(25)\\
    $\Delta_{JK}$ (kHz) & -11.94934(35)& -12.27& & & & $D_{JK}$ (kHz) & -11.78682(38) \\
    $\Delta_{K}$ (kHz) & 92.8838(47) & 93.37& & & & $D_{K}$ (kHz) & 92.7476(50)\\
    $\delta_{J}$ (kHz) & 0.284975(15) & 0.2859 & & & & $d_1$ (kHz) &  -0.285027(16)\\
    $\delta_{K}$ (kHz)  & 4.0972(12) & 3.90 & & &  & $d_2$ (kHz) & -0.0135522(41)\\
    \\
    $\Phi_{J}$ (Hz) & 0.0009617(32) & 0.00144 & & & & $H_J$ (Hz) & 0.0010340(33)\\
    $\Phi_{JK}$ (Hz) & -0.005335(44)& & & & & $H_{JK}$ (Hz) &  -0.006741(47)\\
    $\Phi_{KJ}$ (Hz) &  -0.21085(80)& -0.226 & & & & $H_{KJ}$ (Hz) & -0.20724(84)\\
    $\Phi_{K}$ (Hz) &  -0.2256(33)& & & & & $H_K$ (Hz) &  -0.2286(34)\\
    $\phi_{J}$ (Hz) & 0.0001915(25)& & & & & $h_1$ (Hz) & 0.0001947(26)\\
    $\phi_{JK}$ (Hz) & -0.03251(10) & & & & & $h_2$ (Hz) & -0.00003012(35)\\ 
    \\
    $L_{J}$ (mHz) & -0.00000719(14)& & & & &  $L_{J}$ (mHz) & -0.00000814(15)\\
    $L_{JJK}$ (mHz) & 0.0004567(16)& & & & & $L_{JJK}$ (mHz) & 0.0004312(17)\\
    $L_{JK}$ (mHz) & -0.006856(85) & & & & & $L_{JK}$ (mHz) &  -0.006782(90)\\
    $L_{KKJ}$ (mHz) & 0.03325(57)& & & & & $L_{KKJ}$ (mHz) & 0.03337(60)\\
    $l_{J}$ (mHz) & -0.00000208(12)& & & & &  $l_{1}$ (mHz) & -0.00000233(13)\\
    \\
    $P_{KJ}$ ($\mu$Hz) & -0.001196(58)& & & & & $P_{KJ}$ ($\mu$Hz) & -0.001211(61)\\
    \\
    $N\textsubscript{lines}$\tnote{b}  & 2172& 122 & 38  & & & $N\textsubscript{lines}$\tnote{b} & 2172 \\
    $\sigma\textsubscript{fit}$ (MHz)  & 0.0386& &  & & &  $\sigma\textsubscript{fit}$ (MHz) & 0.0406  \\

    \hline
    \end{tabular}
    \begin{tablenotes}
    \item[a] $\omega$B97X-D/6-31+G(d).
    \item[b] Number of distinct frequencies.
    \end{tablenotes}
    \end{threeparttable}
\end{table*}

The $A$, $B$, and $C$ rotational constants computed from our work match very closely to those derived by \citet{caminati} and \citet{buckley}. In fact, they agree within less than 0.001\% in each case. The constants are also predicted with fairly high accuracy using computational methods, as the  $A$, $B$, and $C$ constants are over-predicted by only 0.13\%, 0.46\% and 0.34\%, respectively.

On the other hand, the distortion constants from this work differ much more notably from those determined by \citet{caminati}. The quartic distortion constants deviate by 0.32\,\%--5.1\,\%. The sextic constants disagree significantly by 6.7\,\% ($\Phi_{KJ}$) and 33.2\,\% ($\Phi_{J}$). This, however, is not entirely surprising since the current work includes transitions at much higher frequencies as well as nine additional higher order distortion terms.

The fits using the A and S reduction were ultimately of similar quality. Both fits include 20 constants. The A reduction resulted in a reduced RMS error but the correlation between the individual constants was slightly greater, indicating that they are not as well determined. Generally, higher order terms were removed unless their inclusion resulted in at least around a 10\,\% improvement of the RMS error of the fit. 

A-E splitting due to the methyl group rotation began to appear in some of the b-type transitions with low-to-mid $K_a^{\prime\prime}$ values. Many of these mid $K_a^{\prime\prime}$ transitions (i.e. $K_a^{\prime\prime} = 9-11$) are observed to be split by approximately 1.1 MHz. That said, none of the observed fairly high $K_a^{\prime\prime}$ Q-branch $\mu_b$ lines displayed any splitting patterns. As of now, the split lines were not included in the fit, and their treatment along with the determination of the methyl rotor parameters will be the subject of a future study. Additionally, the $K_a^{\prime\prime}$ $\geq$ 29 a-type transitions began to deviate notably from the fit above $J \approx 70$ and were thus mostly omitted from the fit. They are likely being perturbed due to interactions with excited vibrational states. This coupling will also be analyzed and treated in future works. Transitions from several of the low-lying vibrationally excited states mentioned in Section~\ref{sec:calc} are visible in the spectrum, some with intensities around half of the ground state. The assignment and fitting of these states will also be included in future works. Finally, while no transitions corresponding to the higher energy conformers are immediately obvious, a more robust investigation will also be conducted.

\section{Observations}

\label{sec:obs}
\subsection{NGC 6334I Observations}
\label{subsec:ngc_obs}
The Band 4 data used in this paper are from the Cycle 5 ALMA program 2017.1.00661.S with (unprojected) baseline lengths ranging from 0.15 - 5.2 km. The data are comprised of four datasets obtained on 2017-12-03 (1), 2018-12-07 (1), and 2018-01-04 (2). The data contain four spectral windows, each with a bandwidth of 937.500 MHz, central (TOPO) frequencies of 130.50, 131.55, 144.47, 145.40 GHz, respectively, and a spectral resolution of 1.1 km s$^{-1}$ corresponding to a channel width of 0.49 MHz.

The data were initially calibrated using the Cycle 6 ALMA Pipeline. Then line-free channels were identified and used to improve the default pipeline uv-continuum subtraction. The same channels were then used to create a continuum image, that was used as the starting point for an iterative self-calibration, the solutions of which were also applied to the continuum-subtracted uv-data prior to making the final image cubes (see e.g. \citet{Brogan18self}).
The native robust=0.5 angular resolution of the resulting continuum image and spectral line cubes was 0$^{\prime\prime}$.23$\times$ 0$^{\prime\prime}$.16. The images were subsequently convolved to a common angular resolution of 0$^{\prime\prime}$.26$\times$ 0$^{\prime\prime}$.26, and primary beam correction was applied before beginning our analysis.

\subsection{IRAS 16293-2422B Observations}
\label{subsec:iras_obs}
We also searched for methoxyethanol toward the B component of the prototostellar system IRAS~16293-2422 using observational data from the ALMA Protostellar Interferometry Line Survey, PILS (\citet{jor16}; project id: 2013.1.00278.S). Methoxymethanol was previously detected toward this source in data from the PILS survey \citep{eme_iras}. PILS consists of an unbiased line survey of IRAS~16293-2422 using ALMA covering frequencies from 329.1 to 362.9~GHz with an angular resolution of 0.5$''$ and spectral resolution of $\approx$0.2~km~s$^{-1}$. We refer to \citet{jor16} for further details about the survey and the data from it.

\section{Observational Analysis and Results}
\label{sec:obs_results}

\begin{figure*}[htb!]    
    \centering
    \includegraphics[width=0.8\linewidth]{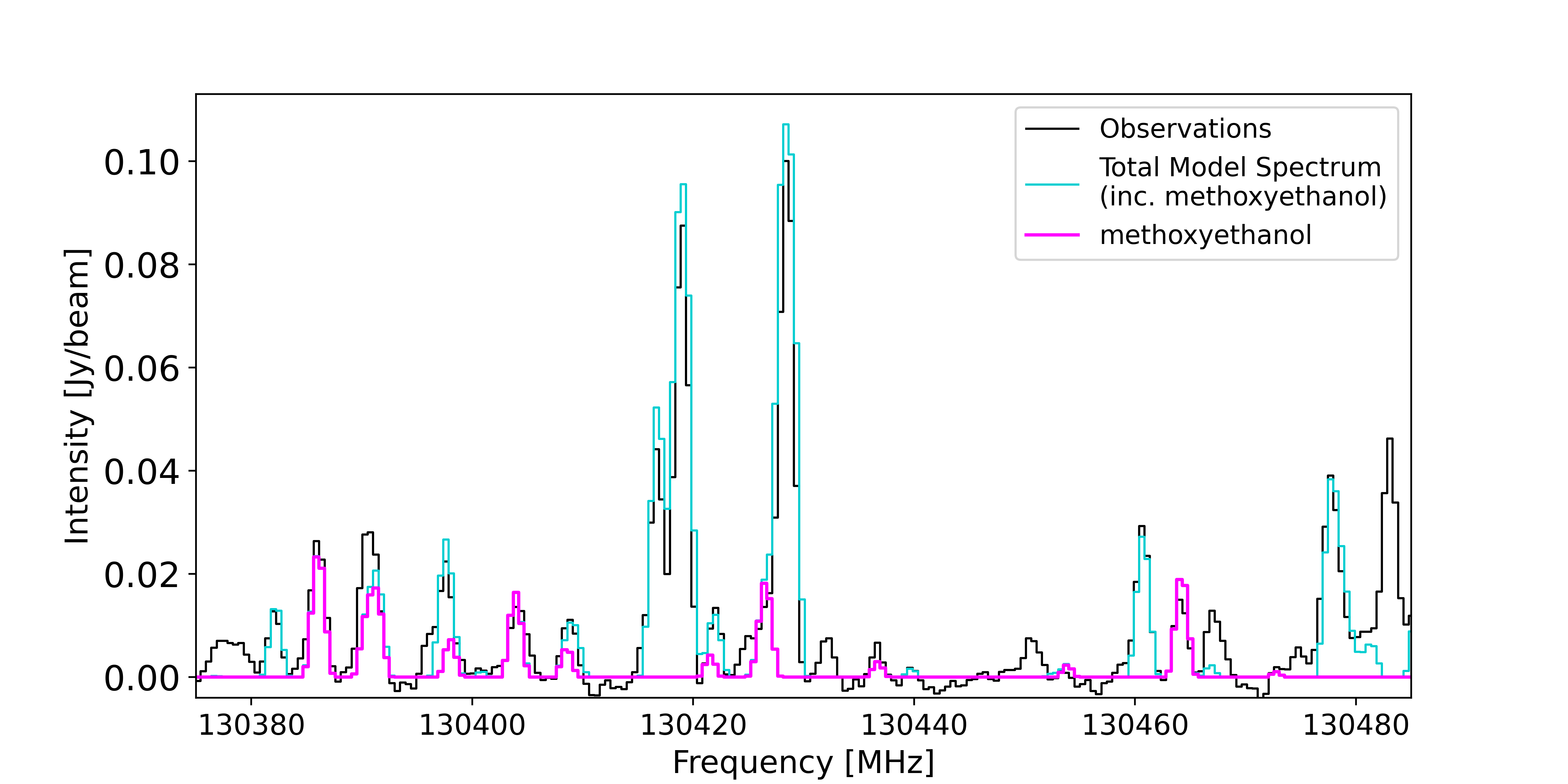}
    \includegraphics[width=0.8\linewidth]{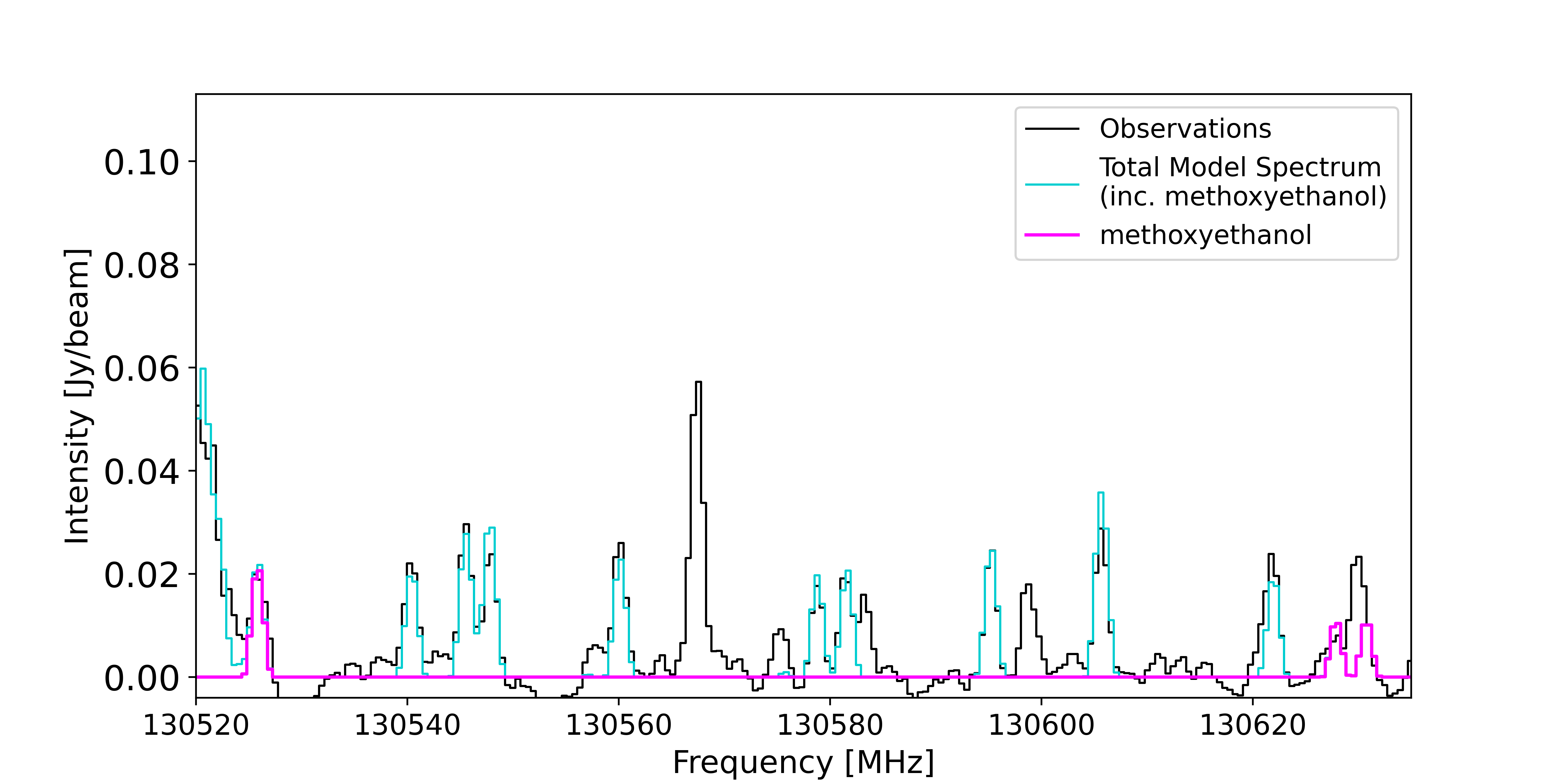}
    \includegraphics[width=0.8\linewidth]{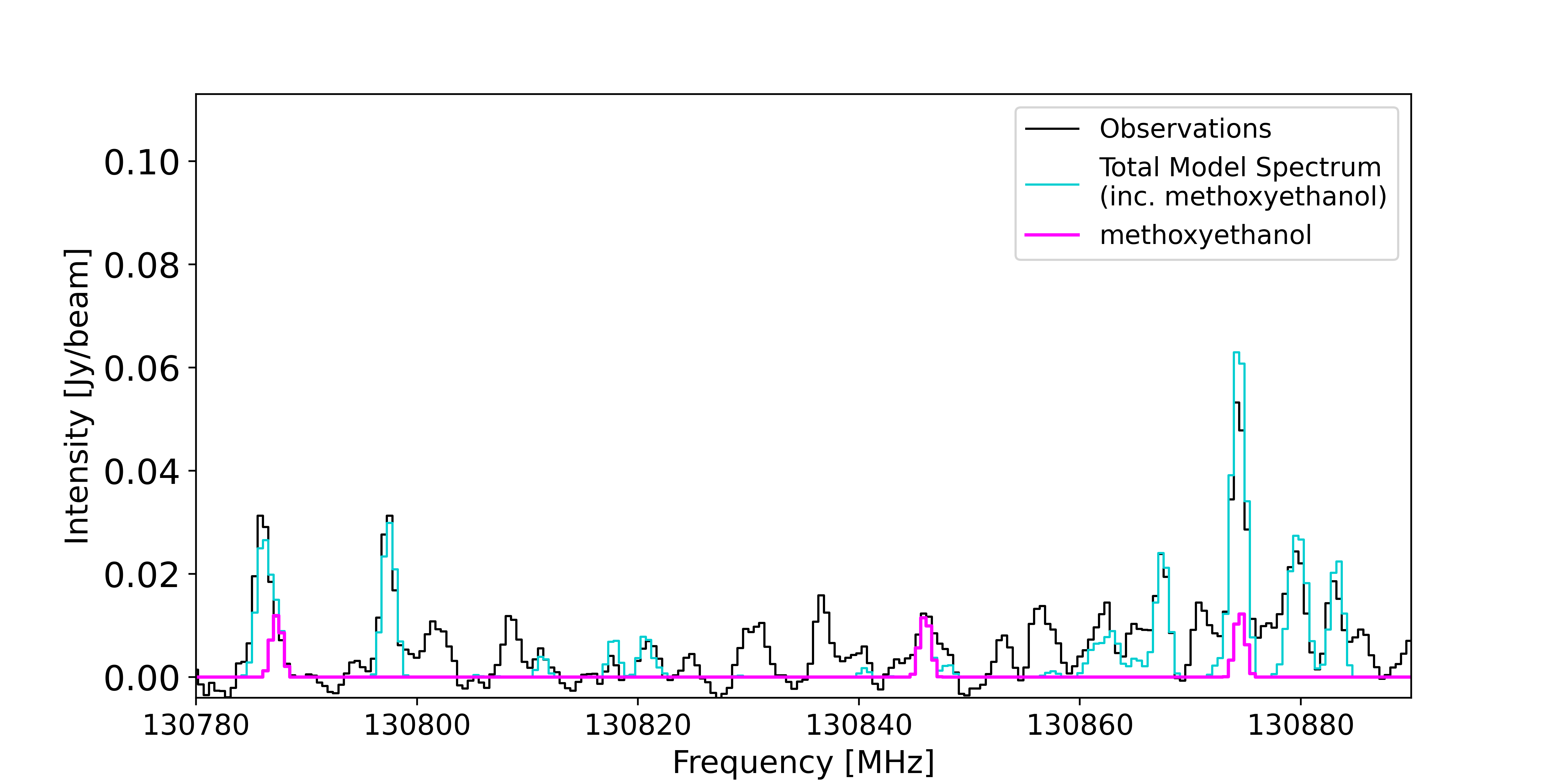}
    \caption{Three frequency windows showing the most prominent 2-methoxyethanol transitions (magenta) in our data toward NGC 6334I. All of the transitions in our frequency range are present in the data, including several that are unblended. The magenta line models 2-methoxyethanol with a column density of $1.3 \times 10^{17}$ cm$^{-2}$ at an excitation temperature of $143$ K.} See Appendix~\ref{apendix_b} for additional spectra.
    \label{spectra_fig}
\end{figure*}

Figure \ref{spectra_fig} shows the detection of 2-methoxyethanol toward the NGC 6334I massive star-forming complex using ALMA Band 4 observations. These data were extracted from the same coordinates that were used to confirm the first interstellar detection of methoxymethanol \citep{methoxy_detection}. A simulated emission spectrum comprised of 20 molecules was used to constrain the physical parameters necessary to model the molecular spectra: these include the excitation temperature, linewidth, velocity, and column density. The source was assumed to fill the beam. The molecular spectra were simulated using the \texttt{molsim} package with a single-excitation temperature model. A decision was made separately to fit all of the molecules with the same linewidth and velocity; this assumption is well-matched to all of the molecules in our model (see Table \ref{mol_table} for a complete list of molecules). The parameters for all of the molecules were simultaneously fit by an automated routine as described in \citet{ngc_fitting}. Overall, we observed 25 methoxyethanol transitions in the Band 4 observations of NGC 6334I. Several of these lines, however, appear to be quite weak in the spectrum. Nine transitions are either completely unblended or only partially blended. All of the observed methoxyethanol transitions are listed in Table~\ref{observed_lines}. The line widths for our frequency range lie between 1.37 and 1.53 MHz. To better explore the uncertainty space for the derived methoxyethanol parameters in particular ($N_T$ and $T_{ex}$), and following the procedures outlined in \citet{Loomis:2018:131}, we performed a Markov Chain Monte Carlo analysis to generate posterior probability distributions for a population diagram analysis for just methoxyethanol \citep{Goldsmith:1999:209}. We conservatively estimate that the flux uncertainty on any given transition is 20\%, which encompasses both the estimated flux calibration uncertainty and a measurement of the noise level of the observations in a nearby line-free region added in quadrature.  Based on these assumptions, we derive final uncertainty estimates of $T_{ex}$=$143_{-39}^{+31}$ K and $N_T$ =  $1.3_{-0.9}^{+1.4} \times 10^{17}$ cm$^{-2}$ for methoxyethanol in this region of NGC 6334I.

\begin{figure}
    \centering
    \includegraphics[width=\columnwidth]{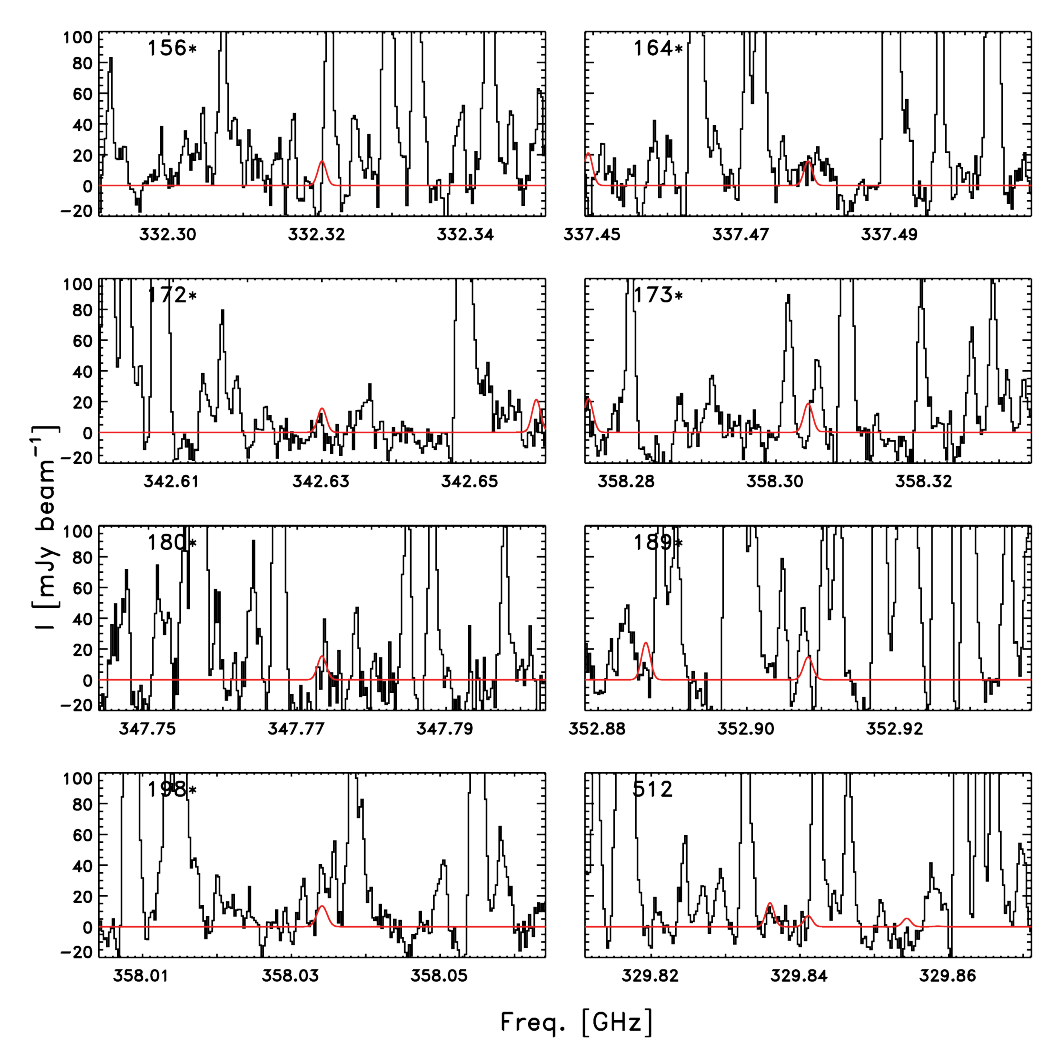}
    \caption{The predicted brightest transitions of 2-methoxyethanol (red) in the frequency range covered by PILS overlaid with the spectrum collected at a position offset 0.5$''$ from the continuum peak of IRAS 16293B (black). No clear molecular transitions of methoxyethanol can be seen. }
    \label{fig:iras_lines}
\end{figure}

For the search presented here toward IRAS 16293B, we analysed the spectrum taken at a position offset by 0.5$''$ from the continuum peak. We calculate synthetic spectra assuming that the gas is in local thermodynamical equilibrium, that the emission can be reproduced with a single column density and excitation temperature, that the emission has a Gaussian distribution with a FWHM of 0.5$''$, and the line width and offset is similar to other species identified in PILS (1~km~s$^{-1}$ FWHM and 2.6~km~s$^{-1}$, respectively). Figure~\ref{fig:iras_lines} shows the eight transitions of methoxyethanol predicted to be the brightest in the frequency range of PILS for an excitation temperature of 100 to 150~K (the excitation temperature of methoxymethanol was found to be 130~K; \citet{eme_iras}). As shown in the figure, there are no clean lines that can be used to claim a detection of methoxyethanol for this source. The plotted synthetic spectrum corresponds to a column density of $2.5\times10^{15}$~cm$^{-2}$, which represents the upper limit to the methoxyethanol column density toward this position. Searching for methoxyethanol at an excitation temperature of 300 K also does not reveal any detected lines, and the upper limit to the column density is even more stringent (lower) than what was found above.

\section{Discussions}
\label{sec:discussion}

\subsection{Possible Formation Pathways}
\label{sub:formation}
\label{subsection:formation}

\subsubsection{Radical-Radical Recombination}
\label{subsub:radical}

Similar to methoxymethanol \citep{methoxy_detection}, the formation of 2-methoxyethanol via radical-radical recombination may stem from reactions containing the methoxy (\ce{CH3O}) and hydroxymethyl (\ce{CH2OH}) radicals. These species can be formed from abundant known interstellar species, such as hydrogen abstraction from methanol in interstellar ices \citep{radical_ice} or in the gas phase \citep{radical_gas}, as well as hydrogen addition to formaldehyde on grains \citep{desorption4}.

Firstly, the 2-hydroxyethyl (\ce{CH2CH2OH}) radical could feasibly react with the methoxy radical to form 2-methoxyethanol.  

\begin{equation} \label{eq:radical_recomb}
        \ce{CH3O + CH2CH2OH -> CH3OCH2CH2OH}
\end{equation}
    
The 2-hydroxyethyl radical may potentially be produced via photodissociation of ethanol. However, previous studies have shown that the predominant H atom product channel of ethanol ultraviolet (UV) photodissociation at various different wavelengths is the breaking of the OH bond to form the ethoxy radical (\ce{CH3CH2O}) \citep{ethanol_photo, ethanol_157}. Another formation route of this radical is the reaction of OH and ethylene (\ce{C2H4}). In fact, the addition of the OH radical into the $\pi$ orbital of the C=C bond of ethylene has been shown to have no overall energetic barrier \citep{hydroxyethyl}.
    
\begin{equation} \label{eq:hydroxyethyl}
        \ce{C2H4 + OH -> CH2CH2OH}
\end{equation}
Once formed through this route, the 2-hydroxyethyl radical is submerged in a notable $>$100 kJ/mol potential well. As demonstrated by \citet{hydroxyethyl}, this radical may therefore be quite efficiently produced via the gaseous radiative association of \ce{OH} and \ce{C2H4} in colder interstellar conditions.

Alternatively, 2-methoxyethanol could also be formed via the radical-radical recombination reaction of \ce{CH2OH} and the methoxymethyl radical (\ce{CH3OCH2}). 

\begin{equation} \label{eq:rad_1}
        \ce{CH2OH + CH3OCH2 -> CH3OCH2CH2OH}
\end{equation}

While neither of these radical species have been previously observed in the interstellar medium, they have both been proposed to form via astrochemically-relevant reactions. For example, along with the aforementioned reactions, \ce{CH2OH} has been observed from UV photolysis of methanol ices \citep{oberg_ice}. Additionally, \ce{CH3OCH2} has been suggested to form via reactions of methylene (\ce{CH2}) and \ce{CH3O} in ices \citep{g22, cocoa} along with hydrogen abstraction from dimethyl ether (\ce{CH3OCH3}) through reactions with OH \citep{shannon} or 
certain gas-phase halogen atoms \citep{balucani}. 

These radical recombination reactions are most likely to occur in interstellar ices where the grain acts as a third body that can absorb excess energy and stabilize the product of the association \citep{herbst_synthetic}. However, similar to the radiative association reactions proposed by \citet{tennis} and \citet{balucani} to produce dimethyl ether, 2-methoxyethanol might also be able to form via radiative association of \ce{CH3O} and \ce{CH2CH2OH} in the gas. That being said, if possible, this process would likely require cold environments, and the effectiveness of this proposed gas-phase pathway is still not fully understood and requires further study (i.e. analysis of the energetics and reaction channels).


\subsubsection{Insertion Reactions}
\label{subsub:insertion}
Along with these radical-radical association reactions, 2-methoxyethanol may also form through oxygen and carbene (\ce{CH2}) insertions into the bonds of smaller molecules on interstellar ices.

Singlet carbene can be formed from interstellar methane through interactions with cosmic rays or energetic electrons \citep{kaiser98, kaiser_ch2}. \citet{kaiser_ch2} have shown that the addition of electronically excited singlet carbene into the various bonds of methanol can form ethanol and dimethyl ether in methane/methanol ice mixtures following interaction with energetic electrons. \citet{bergantini_eme} also propose carbene insertion into the C-H and C-O bonds of dimethyl ether or the O-H bond of ethanol to be possible barrierless formation routes of ethyl methyl ether in irradiated ices. Therefore, a similar carbene insertion reaction into the C-O bond of methoxymethanol or the O-H bond of ethylene glycol can feasibly form 2-methoxyethanol on interstellar ices. 

\begin{equation} \label{eq:carbene1}
        \ce{CH3OCH2OH + CH2 ($a^1A_1$) -> CH3OCH2CH2OH}
\end{equation}

\begin{equation} \label{eq:carbene2}
        \ce{OHCH2CH2OH + CH2 ($a^1A_1$) -> CH3OCH2CH2OH}
\end{equation}

However, the aforementioned study of ethanol and dimethyl ether formation also noted that radical-radical recombination reactions are favored over the carbene addition pathways. 


Another potential pathway to form 2-methoxyethanol is the insertion of electronically excited singlet oxygen atoms, O(\textsuperscript{1}D), into a C-H bond of ethyl methyl ether, which has been previously detected toward several interstellar sources \citep{eme_orion, eme_iras}.

\begin{equation} \label{eq:rad_2}
        \ce{CH3OCH2CH3 + O (\textsuperscript{1}D) -> CH3OCH2CH2OH}
\end{equation}

O(\textsuperscript{1}D) has been shown to form via photolysis of several oxygen bearing molecules, such as water, carbon dioxide, and \ce{O2}, which are highly abundant in interstellar ices \citep{o1,o2,o3}. In a theoretical study of reactions of O(\textsuperscript{1}D) with methanol, dimethyl ether and methyl amine, \citet{hays13} showed that the O(\textsuperscript{1}D) insertion pathways into the C-H bonds are barrierless and exothermic. Additionally, experiments have indicated that the direct O(\textsuperscript{1}D) insertion into ethane and methane can form ethanol and methanol with a negligible energetic barrier \citep{bergner19,bergner17}. Therefore, a similar O(\textsuperscript{1}D) insertion pathway could also likely occur with ethyl methyl ether to form 2-methoxyethanol. We expect these oxygen insertion reactions to occur on interstellar ices where the direct products can be collisionally stabilized instead of further decomposing \citep{bergner19}.


\subsection{Abundance Differences Between Sources}
\label{sub:difference}

The column density of 2-methoxyethanol in NGC 6334I and the upper limit derived toward IRAS 16293B vary by several orders of magnitude. However, this is mainly representative of different number densities of the gas in these two regions. A more informative metric to gauge the relative underproduction of 2-methoxyethanol in IRAS 16293B may therefore be the methoxymethanol/methoxyethanol ratios in these sources. In NGC 6334I, this ratio is $\sim$31 \citep{methoxy_detection}, while it is $>56$ in IRAS 16293B \citep{eme_iras}. This difference may suggest that methoxyethanol is more efficiently produced in NGC 6334I. Alternatively, if there is a notable difference between the binding energies of methoxyethanol and methoxymethanol, the abundance difference may also stem from more effective desorption of this species from the ice in NGC 6334I.


Several groups have previously noted differentiation of relative molecular abundances between NGC 6334I and IRAS 16293B, including \ce{NH2CN}/\ce{NH2CHO} \citep{amide_ngc},  \ce{CH3OCHO}/\ce{OHCH2CHO} \citep{mf_ga}, and \ce{CH3NH2}/\ce{CH3OH} \citep{methylamine}. It is therefore clear that the different physical environments of these two sources may have an impact on the chemistry that occurs. 

Because both of these sources contain embedded protostars, radiation may be notably effecting the chemistry. Previous studies of these sources hypothesize that UV radiation from the central protostar or variations in the strength of the radiation field may impact the observed chemical abundances (e.g. \citealt{jor18,amide_ngc}). In fact, many of the molecules involved in the proposed reactions in Section~\ref{sub:formation} are produced via interaction of precursor species with radiation. For example, the \ce{CH2OH} and \ce{CH3O} radicals can be formed via photodissociation of methanol \citep{oberg_ice,laas_ch3o}. Also, as mentioned previously, singlet oxygen can be generated via photolysis of various oxygen containing molecules. Therefore, the potentially enhanced radiation field of the high mass star forming region NGC 6334I (compared to the low mass source IRAS 16293B) may promote the production of the required precursor species to methoxyethanol.

 As was proposed by \citet{methoxy_detection}, the addition of O($^1$D) to dimethyl ether (DME) may form interstellar methoxymethanol. Therefore, the DME/methoxymethanol ratios in these sources can potentially suggest how efficient the oxygen insertion reactions are. In IRAS 16293B, the DME/methoxymethanol ratio is 1.7 \citep{jor18,eme_iras}. However, this ratio is 0.35 toward the position of NGC 6334I investigated in this work. While it is likely that there are various other physical and chemical factors contributing to this result, it could suggest that oxygen insertion processes are more efficient in NGC 6334I, possibly due to a greater abundance of electronically excited oxygen on the interstellar ices.

Furthermore, the abundance of the methoxy radical (or simply the efficiency of methoxylation) in these sources can be indirectly gauged from the ethanol/methoxyethanol ratios. In NGC 6334I, this value is approximately 12 (with the derived ethanol column density being $1.59 \times 10^{18}$ cm$^{-2}$). However, in IRAS 16293B, the ratio is $>92$ \citep{jor18}. This may, in part, suggest that the methoxy radical is more favorably produced in NGC 6334I, and therefore is more available for reaction with other organic radicals.

Additionally, as was noted by \citet{methylamine} in their discussion of the under-abundance of \ce{CH3NH2} in IRAS 16293B compared to NGC 6334I, warmer dust temperatures during the time of methanol formation in NGC 6334I may have permitted greater radical mobility on grain surfaces. Therefore, some of the larger radicals that are proposed to recombine to form 2-methoxyethanol might have been able to more efficiently traverse the grains, thus increasing their likelihood of interaction.

Finally, the variation in the methoxymethanol to methoxyethanol ratios in these two sources may be due to different binding energies of the two species. Several groups have investigated the temperature programmed desorption (TPD) of methoxymethanol from irradiated methanol ices. They report m/z = 61 desorption features corresponding to methoxymethanol ranging from $\sim$155-170 K \citep{desorption1, desorption2, desorption3, desorption4}. That being said, to our knowledge no similar studies have reported the desorption temperature or binding energy of methoxyethanol. If methoxyethanol has a notably larger binding energy than methoxymethanol, desorption differences may result in the lack of gaseous methoxyethanol in IRAS 16293B. In fact, as was highlighted by \citet{amide_ngc}, a compact low-mass protostellar source may have a more rapidly decreasing radial temperature gradient. On the other hand, a high mass star forming region is likely to adequately heat larger regions of space. Consequently, it is possible that methoxyethanol is not able to desorb over a larger percent of the observed region in IRAS 16293B, leading to a lower gas-phase abundance. However, it has been seen with other complex organic molecules that the TPD profiles from water-ice mixtures are strongly controlled by the desorption of water \citep{water_desorption}. Therefore, if these species are trapped in water ice, the altered desorption mechanisms may result in the desorption temperatures being similar for these two molecules. It is clear that further experiments of the desorption mechanisms and temperatures of these molecules are required to draw any definitive conclusions.

\subsection{Validation of Machine Learning Prediction}
\label{sub:ml}
The Bayesian ridge and Gaussian process regressors trained by \citet{fried} both overpredicted methoxyethanol to have a column density of $\sim$$10^{17}$ cm$^{-2}$ toward IRAS 16293B. This overprediction is likely in part because the most chemically similar training examples are all smaller, less complex, and more abundant than methoxyethanol. This molecule is larger than practically all other species in the detected dataset. 3-hydroxypropenal is the only species in the training data that contains as many non-hydrogen atoms as 2-methoxyethanol, and the detection of 3-hydroxypropenal in IRAS 16293B was only tentative at the time of this work \citep{hydroxypropenal}. This model, however, bases its predictions on relationships in chemical vector space. The detected molecule with the closest chemical vector representation to methoxyethanol is methoxymethanol, and, as mentioned previously, there is very little training data for similar molecules of comparable size/complexity. Consequently, the model is presumably over-fitting to the methoxymethanol training example and predicting methoxyethanol's column density to be nearby that of methoxymethanol. Despite the numerical inaccuracy of this prediction, the model can still provide molecular candidates that are nearby in chemical vector space to the detected species in the source. Therefore, the other molecules predicted by the model, such as methanediol (\ce{OHCH2OH}), may still be strong candidates for detection due to their chemical similarity to the known molecular inventory. Now, with these new results, devising a method to incorporate upper limits into the model training, and thus further expanding the training set, could likely allow for greater refinement of the model, less over-fitting, and a decrease in the column density overpredictions.

\section{Conclusions}
\label{sec:conclusion}

In this work, we have collected and analyzed the rotational spectrum of 2-methoxyethanol (\ce{CH3OCH2CH2OH}) in the frequency regions of $8.7-26.5\,$GHz, $68-85 \,$GHz, and $150-500\,$GHz. Using these data, we report the first interstellar detection of this molecule using Band 4 ALMA observations toward the massive protocluster NGC 6334I. 25 molecular transitions are identified, nine of which are either fully unblended or only partially blended. In all, a column density of $1.3_{-0.9}^{+1.4} \times 10^{17}$ cm$^{-2}$ was derived at an excitation temperature of $143_{-39}^{+31}$ K. This molecule was also searched for in Band 7 ALMA observations toward source B of the low-mass protostellar system IRAS 16293. However, clear molecular transitions were not observed and an upper limit column density of $2.5 \times 10^{15}$\,cm$^{-2}$ was determined. We proceeded to discuss potential formation pathways of the molecule, especially focusing on those stemming from known interstellar precursors. Among the plausible formation routes are the additions of electronically excited carbene into methoxymethanol and ethylene glycol or the insertion of excited singlet oxygen atoms into ethyl methyl ether. This molecule may also be formed through the recombination of various astrochemically-relevant radical species, including \ce{CH3O} and \ce{CH2CH2OH}. 

We also discuss the causes of the relative methoxyethanol under-abundance in IRAS 16293B compared to NGC 6334I. These include the potentially enhanced radiation field of a high mass star forming region like NGC 6334I resulting in a greater production of the methoxyethanol precursors. Additionally, warmer dust temperatures during the earlier stages of star formation in NGC 6334I may have allowed for greater radical mobility on interstellar dust grains, and thus enabled the fairly large radical fragments that are proposed to form methoxyethanol to traverse the grains and recombine.

With this detection of methoxyethanol in NGC 6334I, the methoxylated counterparts of methanol, ethanol, methane, and formaldehyde have all been observed toward this star-forming region. The methoxy radical therefore likely is an important factor in the chemistry of this source and it may be a strong candidate for the detection of other species containing the methoxy functional group. Further detection and analysis of methoxylated molecules would allow for a better constraint of the chemistry involving the methoxy radical, which may in turn provide information about the dissociation of methanol during the process of star formation.

\begin{acknowledgments}
We would like to sincerely thank the four anonymous referees who each provided expert criticisms which substantially improved the quality of this manuscript. ZTPF and BAM gratefully acknowledge the support of Schmidt Family Futures. GW and BAM gratefully acknowledge the support of the Arnold and Mabel Beckman Foundation Beckman Young Investigator Award. BAM gratefully acknowledges support of National Science Foundation grant AST-2205126. The authors gratefully acknowledge Dr. Ceci Xue for helpful discussions regarding potential formation pathways of this molecule. ZTPF, BAM, and GW gratefully acknowledge Professor Robert W. Field for providing equipment for the construction of the CP-FT mm-wave instrument.  The National Radio Astronomy Observatory is a facility of the National Science Foundation operated under cooperative agreement by Associated Universities, Inc. This research received computational support from the PIA ANR project CaPPA (ANR-11-LABX-0005-01). This paper makes use of the following ALMA data: ADS/JAO.ALMA\#2017.1.00661.S, ADS/JAO.ALMA\#2013.1.00278.S, and ADS/JAO.ALMA\#2012.1.00712.S. ALMA is a partnership of ESO (representing its member states), NSF (USA) and NINS (Japan), together with NRC (Canada), MOST and ASIAA (Taiwan), and KASI (Republic of Korea), in cooperation with the Republic of Chile. The Joint ALMA Observatory is operated by ESO, AUI/NRAO and NAOJ.

\textbf{Statement of Efforts.}  All authors contributed to the writing and editing of the manuscript. ZTPF: Spectroscopy, Analysis, and Discussion; SJE: NGC 6334I Analysis; BMH, GW: Spectroscopy and Analysis; ANB: Chemical Pathways and Discussion; LM, RM, STS, MPH: Spectroscopy and Analysis; JJ: IRAS 16293 Analysis; CLB, TRH, AJR, AL, RAL: NGC 6334I Observations and Analysis; BAM: Project Design, NGC 6334I Observations and Analysis.
\end{acknowledgments}


\appendix

\renewcommand\thefigure{\thesection\arabic{figure}}   
\renewcommand\thetable{\thesection\arabic{table}}    

\setcounter{figure}{0}    
\setcounter{table}{0} 

\section{Appendix A: Description of Laboratory Experiments}
\label{apendix_a}

\begin{figure}[h!]
\begin{center}
\includegraphics[width=\columnwidth]{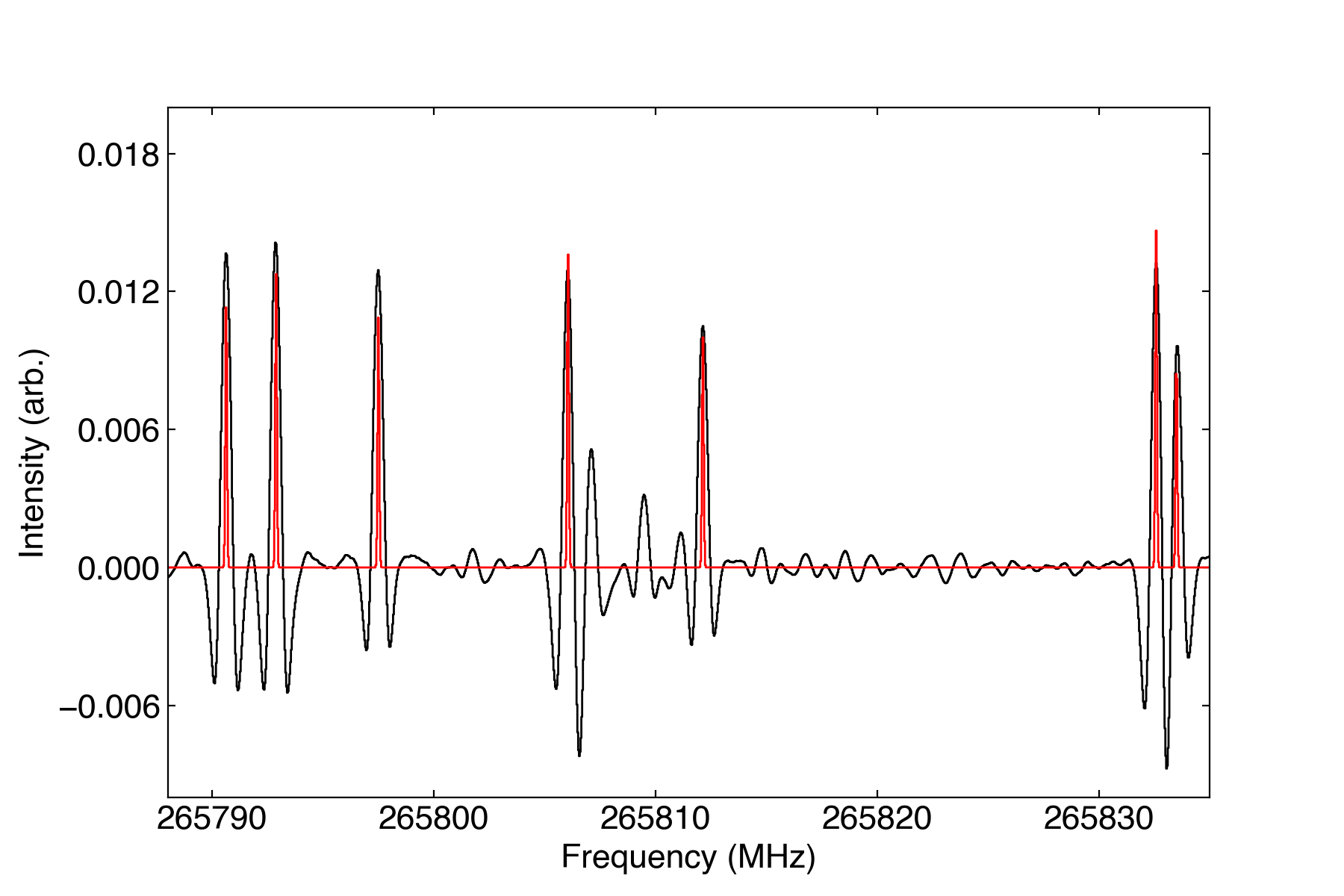}
\caption{
Experimental rotational spectrum of 2-methoxyethanol (black) overlaid with the simulated spectrum of the ground vibrational state of the tGg- conformer from the experimentally fitted parameters (red). The spectral simulation was conducted using SPCAT and the plot was generated with \texttt{molsim} \citep{molsim}. The transitions are various $J = 51$ R-branch $\mu_a$ transitions with $K_a^{\prime\prime}$ values ranging from 15--21. Characteristic $2f$ frequency modulation lineshapes can be seen.}\label{fig:spectrum}
\end{center}
\end{figure}

 The Lille spectrometer with fast-scan mode was used to collect the spectrum in three frequency regions: $150-220\,$GHz, $230-330\,$GHz, and $370-500\,$GHz. This Lille spectrometer was described by \citet{lille}, and the fast-scan setup was detailed by \citet{fast_scan}. In summary, the absorption cell consists of a 220 cm long stainless-steel tube with a 6 cm diameter. The source of radiation was an Agilent synthesizer ($12.5-18.25\,$GHz), and the entire frequency range was covered using combinations of a range of active and passive frequency multipliers. The frequency was modulated at $40.5\,$kHz. The detected output was then demodulated at twice the modulation frequency ($2f$ frequency demodulation), resulting in the characteristic lineshapes shown in Figure~\ref{fig:spectrum}. The spectrum was collected at room temperature. Due to the relative stability of this molecule, the spectrum collection was performed in ``static" mode, in which a fairly consistent quantity of gas (measuring approximately $2-2.5\,$Pa) was contained in the cell. The scanning rate was 1 ms/point, and each spectrum was averaged for 32, 16, or 4 times depending on the frequency range. Due to the Doppler broadening of the spectral lines at higher frequencies, the frequency step was increased from $30-54\,$kHz as the experiment progressed. The uncertainties of the line frequencies were estimated to be 50 kHz throughout the entire frequency range.

We collected additional mm-wave data from $68-85\,$GHz using a newly updated chirped-pulse Fourier transform (CP-FT) mm-wave spectrometer at MIT. The design of this spectrometer was largely inspired by those described by \citet{park} and \citet{barnum}. In this instrument, a chirped pulse is first generated using a $12.5\,$GS/s arbitrary waveform generator (AWG) that is capable of producing chirps from $0-5\,$GHz. This chirped pulse is then mixed with a $9.375\,$GHz signal that is produced by a phase-locked dielectric resonator oscillator (PDRO) which is phase-locked to a 10 MHz Rb frequency standard. The upper or lower sideband can then be selected using different bandpass filters. The resulting frequencies are amplified and multiplied by a factor of eight using two multipliers. Depending on whether the upper or lower sideband is selected, the spectrometer operates in the up-converted frequency ranges of $68-74.4\,$GHz or $79.2-94.4\,$GHz. The signal is emitted via a rectangular gain horn that is positioned outside of the chamber. The beam is collimated and focused using polytetrafluoroethylene (PTFE) lenses and enters the vacuum chamber through a PTFE window. The molecular sample is seeded in a carrier gas and is introduced into the chamber via a pulsed supersonic expansion at a 4 Hz repetition rate. The pulsed molecular beam interacts with the chirped-pulse in a perpendicular orientation. The spectrometer operates in two modes, either probing each molecular gas pulse with only one chirped pulse, or with 10 chirps interacting with each individual gas pulse yielding an effective $40\,$Hz repetition rate. Following the interaction of the radiation with the molecular sample, the molecular free induction decay (FID) exits the chamber through a PTFE window and is collected using PTFE lenses in combination with a receiving rectangular gain horn. For the down-conversion of the FID, the local oscillator (LO) is generated by mixing the $9.375\,$GHz signal from the PDRO with a single frequency signal produced by an additional channel of the AWG. This LO is then multiplied by a factor of six (to reach $75\,$GHz) and mixed with the FID in order to down-convert it into the microwave region. This signal is finally digitized by a Keysight $80\,$GS/s, $20\,$GHz fast oscilloscope.

In our specific experiment, a cotton ball was soaked with the 2-methoxyethanol liquid sample and placed in a sample holder directly up-stream of the pulsed nozzle. The sample holder was held at a temperature of $\sim$40 $^{\circ}$C. Helium carrier gas with a backing pressure of 2 bar was then flowed through the sample holder and the mixed sample was introduced into the chamber by the pulsed nozzle. In order to optimize the line shape and signal-to-noise ratio, only the first 2 $\mu$s of the FID was Fourier transformed. This was due to the common occurrence of short FID dephasing times of the molecular transitions in the mm-wave region \citep{park}. The transitions were recorded individually using 10 MHz broad chirped pulses centered around each predicted frequency. Using these narrow chirps, a signal-to-noise ratio of $\sim$40 could be achieved by averaging only 3,000 shots in the time-domain. All transitions in this frequency range were estimated to have uncertainties of $50\,$kHz

Finally, data from $8.7-26.5\,$GHz was collected at the New College of Florida using a waveguide-based CP-FT spectrometer. A thorough description of this spectrometer is present in the work of \citet{ncf}. In this setup, a $0.1-4.9\,$GHz linear frequency sweep is first generated with an AWG. It is then filtered, mixed, and amplified before interacting with the molecular sample in a 10 m coiled WRD-750 waveguide. The resulting FID (following amplification, filtering and down-conversion) is digitized by an oscilloscope and Fourier transformed into frequency space. In the specific 2-methoxyethanol experiment, the spectrum was measured in three frequency sections: $8.7-13.5$, $13.5-18.3$, and $18.0-26.5\,$GHz. The sample had a temperature of $253\,$K and a pressure of $5\,$mTorr. The FID collection time was 4 $\mu$s and it was averaged for one million shots in the time-domain. The peak frequencies in the cm-wave data were also estimated to have 50 kHz uncertainties.

\section{Appendix B}
\label{apendix_b}

\subsection{Additional Spectra}
Appendix B.1. contains a figure that displays the additional observed lines of 2-methoxyethanol that were not included in the figures within the main text.

\renewcommand{\thefigure}{B1}
\begin{figure*}[htb!]    
    \centering
    \subfigure[]{\includegraphics[width=0.49\linewidth]{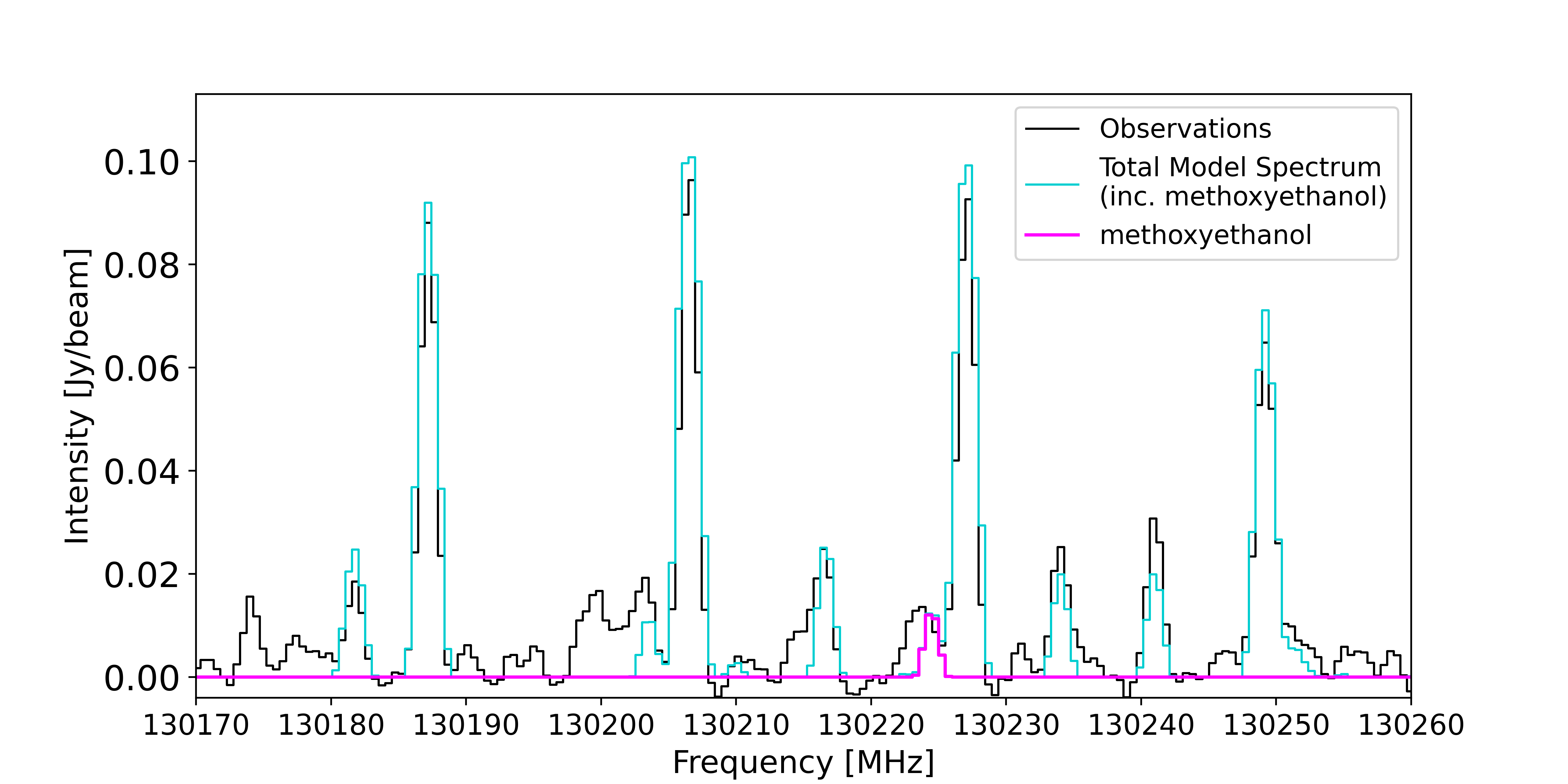}}
    \subfigure[]{\includegraphics[width=0.49\linewidth]{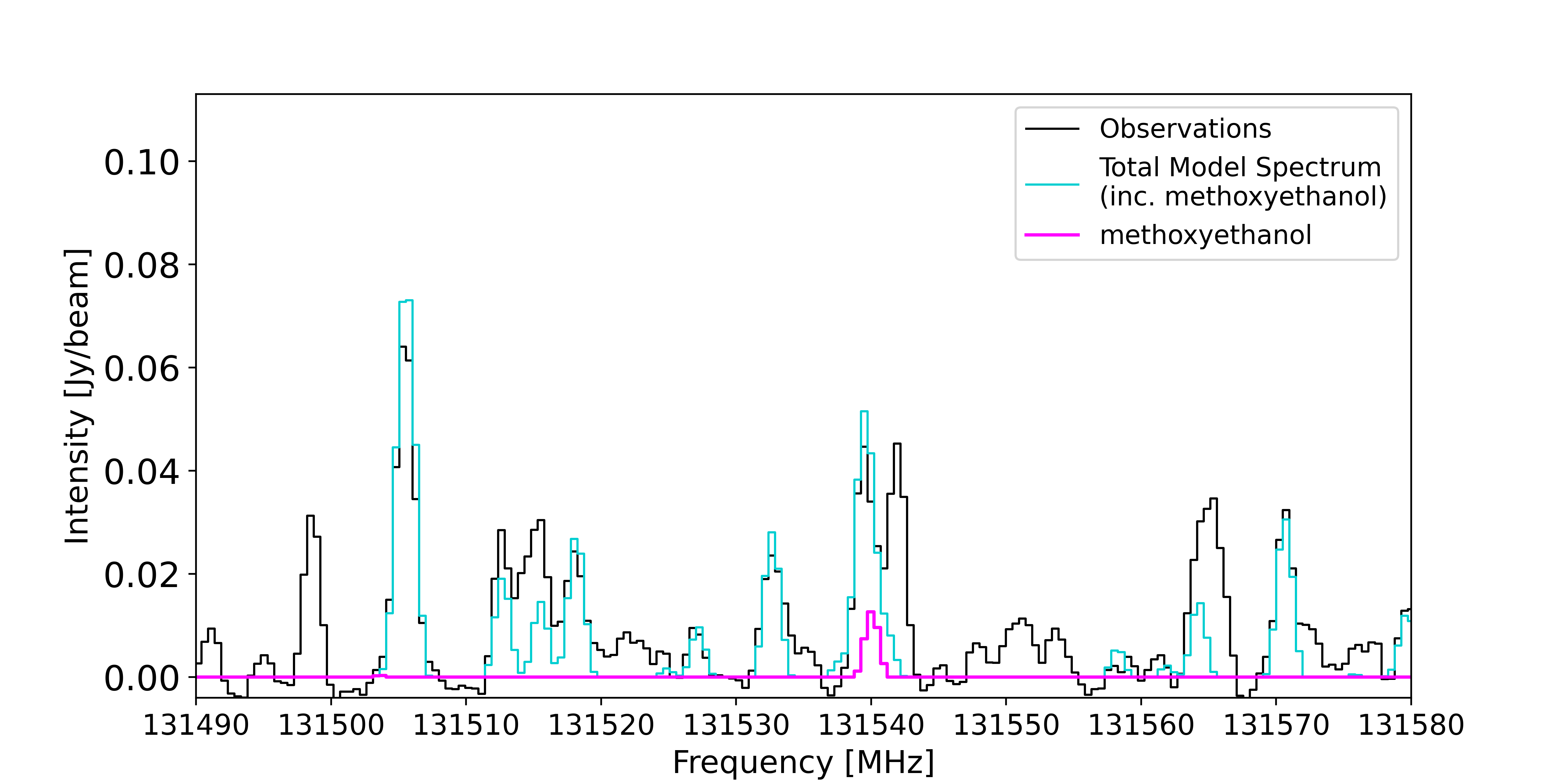}}
    \subfigure[]{\includegraphics[width=0.49\linewidth]{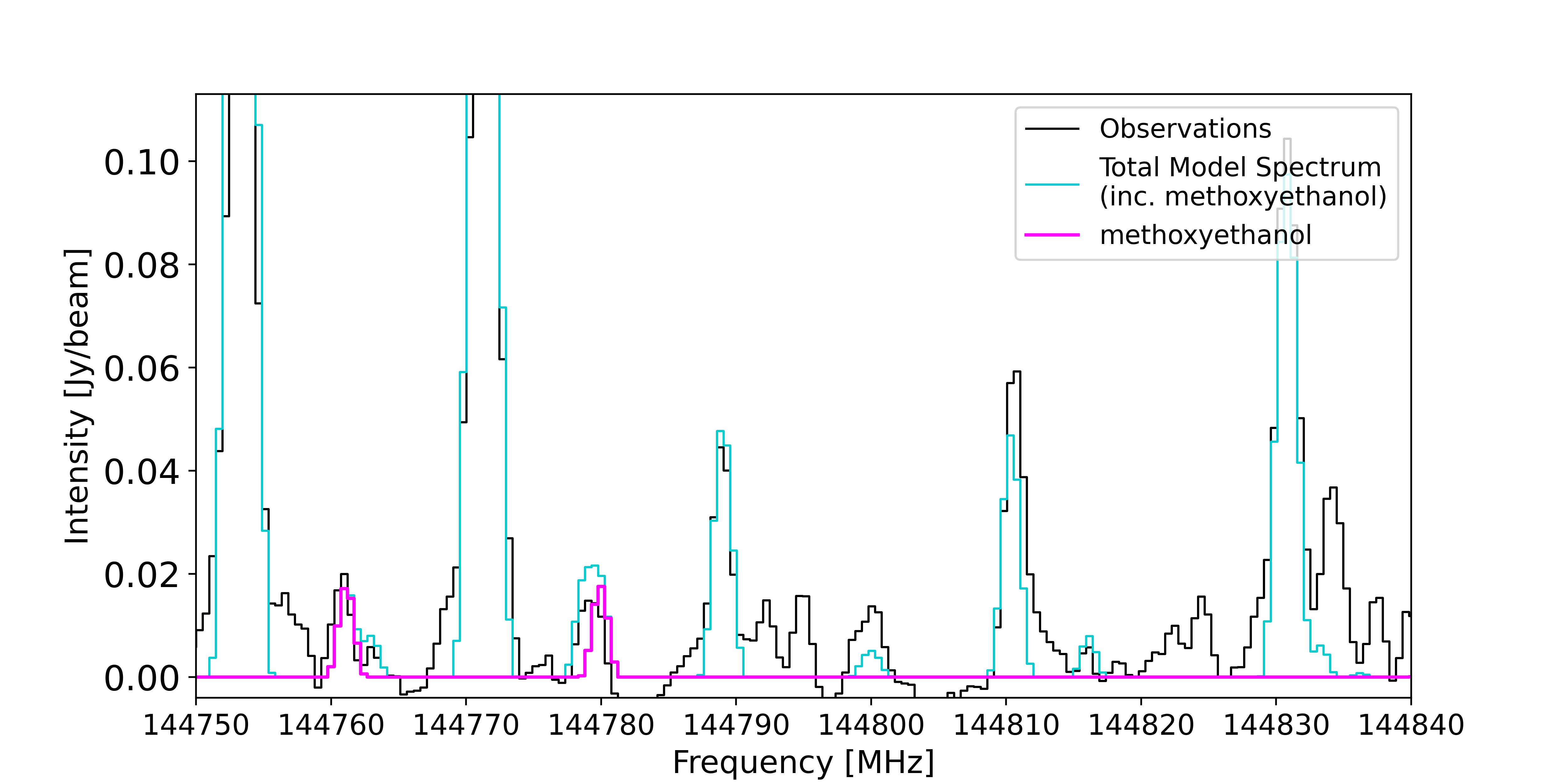}}
    \subfigure[]{\includegraphics[width=0.49\linewidth]{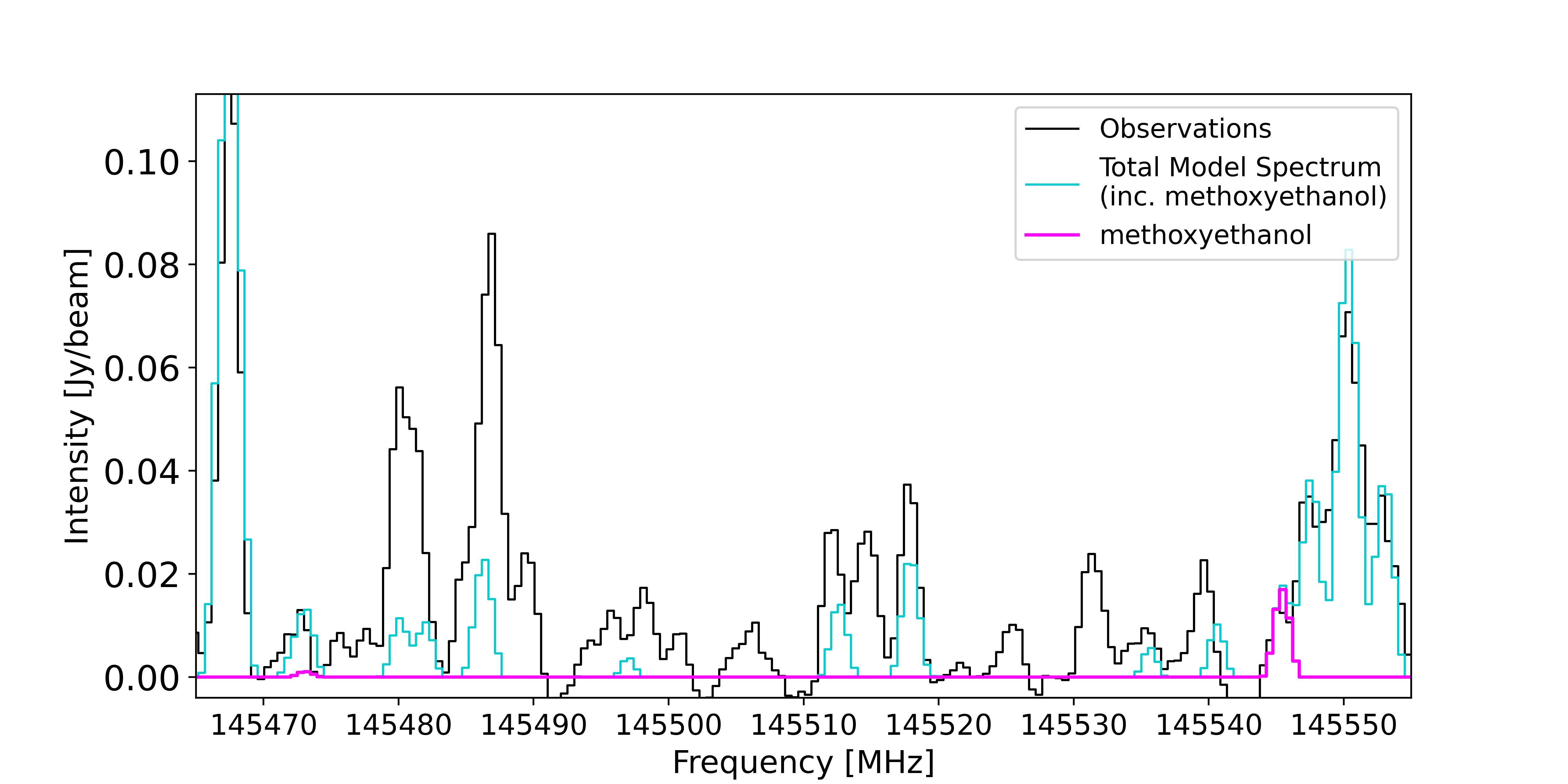}}
    \caption{The remaining detectable 2-methoxyethanol lines in our data range toward NGC 6334I. All of the emission lines are present and accounted for.}
    \label{appendix_spectra_fig}
\end{figure*}

\subsection{Model Molecules, Spectroscopic Information, and Observed Transitions}

Appendix B.2. contains three tables. Table B1 provides the list of molecules included in the simulated model of the NGC 6334I emission. Next, Table B2 notes the observed 2-methoxyethanol transitions in the data toward NGC 6334I. Finally, Table B3 contains information on the partition function of 2-methoxyethanol at various temperatures.

\begin{table*}
\centering
\footnotesize
\caption{Molecules included in the simulated model of NGC 6334I.}
\label{mol_table}
\begin{tabular}{ccc}
\hline

\ce{H2CO} & \ce{CH3OH} & $^{13}$CH$_3$OH \\ 
\ce{NH2CN} & \ce{HNCO} & C$^{33}$S \\
\ce{CH3CHO} & \ce{NH2CHO} & \ce{CH3OCH3}$^a$ \\
\ce{C2H5OH} & \ce{C2H5CN} & \ce{CH3COCH3} \\
\ce{CH3OCHOehto}$^a$ & \ce{HCOCH2OH} & \ce{CH3COOH} \\
\ce{a-(CH2OH)2} & \ce{g-(CH2OH)2} & \ce{CH3OCH2OH} \\
\ce{SO2} & \ce{CH3OCH2CH2OH}$^b$ \\

\hline
\multicolumn{3}{l}{$^a$Catalogs obtained from the JPL spectroscopic database \citep{jpl_database}.} \\
\multicolumn{3}{l}{All other catalogs were obtained from CDMS \citep{cdms_database}.}\\
\multicolumn{3}{l}{$^b$This work.}\\

\end{tabular}
\end{table*}

\begin{table*}
\centering
\caption{Observed transitions of 2-methoxyethanol toward NGC 6334I. It is also noted whether the rotational signal is blended with other molecular lines in the observations. }
\label{observed_lines}
\begin{tabular}{ccccc}
\hline
Rest Frequency (MHz) & Quantum Numbers & $E_{up}$ (K) & $S_{ij}\mu^2$ (D$^2$) & Blended?\\
\hline
130141.4853 & 26 1 26 -- 25 0 25 & 85.16 & 29.975 & Yes \\
130221.7380 & 25 3 23 -- 24 3 22 & 85.88 & 101.496 & Yes \\
130383.3002 & 25 12 13 -- 24 12 12	& 152.96 & 79.321 & No \\
130383.3002 & 25 12 14 -- 24 12 13 & 152.96 & 79.321 & No \\
130383.3495 & 25 13 12 -- 24 13 11 & 165.38 & 75.209 & No \\
130383.3495	& 25 13 13 -- 24 13 12 & 165.38 & 75.209 & No \\
130387.6063 & 25 14 11 -- 24 14 10 & 178.79 & 70.753 & Yes \\
130387.6063 & 25 14 12 -- 24 14 11 & 178.79 & 70.753 & Yes \\
130388.6374 & 25 11 14 -- 24 11 13 & 141.53 & 83.120 & Yes \\
130388.6374 & 25 11 15 -- 24 11 14 & 141.53 & 83.120 & Yes \\
130395.2750 & 25 15 10 -- 24 15 9 & 193.18 & 65.968 & Yes \\
130395.2750 & 25 15 11 -- 24 15 10 & 193.18 & 65.968 & Yes \\
130401.1723 & 25 10 15 -- 24 10 14 & 131.10 & 86.579 & No \\
130401.1723 & 25 10 16 -- 24 10 15 & 131.10 & 86.579 & No \\
130405.8022 & 25 16 9 -- 24 16 8 & 208.56 & 60.860 & Yes \\
130405.8022 & 25 16 10 -- 24 16 9 & 208.56 & 60.860 & Yes \\
130418.7930 & 25 17 8 -- 24 17 7 & 224.93 & 55.415 & Yes \\
130418.7930 & 25 17 9 -- 24 17 8 & 224.93 & 55.415 & Yes \\
130423.8106 & 25 9 16 -- 24 9 15 & 121.65 & 89.720 & Yes \\
130423.8106 & 25 9 17 -- 24 9 16 & 121.65 & 89.720 & Yes \\
130433.9588 & 25 18 7 -- 24 18 6 & 242.27 & 49.641 & Partial \\
130433.9588 & 25 18 8 -- 24 18 7 & 242.27 & 49.641 & Partial \\
130461.4707 & 25 8 18 -- 24 8 17 & 113.21 & 92.514 & No \\
130461.4728 & 25 8 17 -- 24 8 16 & 113.21 & 92.514 & No \\
130523.0445 & 25 7 19 -- 24 7 18 & 105.76 & 95.000 & No \\
130523.1367 & 25 7 18 -- 24 7 17 & 105.76 & 95.000 & No \\
130625.1238 & 25 6 20 -- 24 6 19 & 99.32 & 97.145 & Partial \\
130628.0123 & 25 6 19 -- 24 6 18 & 99.32 & 97.141 & Yes \\
130784.5518 & 25 5 21 -- 24 5 20 & 93.90 & 98.948 & Yes \\
130843.2516 & 25 5 20 -- 24 5 19 & 93.91 & 98.952 & Partial \\
130871.7622 & 25 4 22 -- 24 4 21 & 89.48 & 100.423 & Yes \\
131537.2608 & 25 4 21 -- 24 4 20 & 89.61 & 100.434 & Yes \\
144696.8419 & 29 0 29 -- 28 1 28 & 105.28 & 33.986 & Yes \\
144758.0570 & 29 1 29 -- 28 1 28 & 105.29 & 119.242 & No \\
144776.8686 & 29 0 29 -- 28 0 28 & 105.28 & 119.238 & Yes \\
144838.0837 & 29 1 29 -- 28 0 28 & 105.29 & 33.982 & Yes \\
145542.3426 & 28 3 26 -- 27 3 25 & 106.10 & 113.970 & Partial \\
145619.2973 & 28 2 27 -- 27 1 26 & 102.43 & 23.465 & Yes \\
\hline
\end{tabular}
\end{table*}

\begin{table*}
\centering
\caption{Partition function of 2-methoxyethanol at various temperatures. The rotational partition function was computed using SPCAT. The vibrational partition function was calculated with the harmonic approximation formalism described in \citet{gordy_cooke} using the computed harmonic fundamental vibrational modes. Finally, the total partition function was derived by multiplying the rotational and vibrational partition function values at the respective temperatures. }
\label{rot_q}
\begin{tabular}{cccc}
\hline
Temperature (K) & $Q_{\mbox{rot}} (T)$ & $Q_{\mbox{vib}} (T)$ & $Q_{\mbox{tot}} (T)$ \\
\hline
300.000 & 93781.3044 & 17.8077 & 1670027.8801 \\
225.000 & 60885.5696 & 7.0448 & 428924.5391\\
150.000 & 33112.9351 & 2.8451 & 94209.1014\\
75.000 & 11698.5316 & 1.2989 & 15195.6373\\
37.500 & 4136.1491 & 1.0318 & 4267.8401\\
18.750 & 1463.5092 & 1.0007 & 1464.5855\\
9.375 & 518.4432 & 1.0000 & 518.4435\\
\hline
\end{tabular}
\end{table*}

\end{document}